\documentclass[12pt, letterpaper]{article} 
\usepackage{geometry, natbib, hyperref, graphicx}
\usepackage{amssymb}
\usepackage{verbatim}
\usepackage{subfigure,wrapfig}
\usepackage{booktabs}
\usepackage{enumitem}
\setlist[enumerate]{itemsep=0mm}
\usepackage{float}
\usepackage{lscape}

\def\BE{\begin{equation}}
\def\EE{\end{equation}}

\usepackage[font=small,skip=0pt]{caption}

\newcommand{\eyja}{Eyjafjallaj\"okull}
\title{The development of volcanic ash cloud layers over hours to days
  due to turbulence layering}

\author{Marcus Bursik $^{1,}$*ORCID:{0000-0002-9312-5202}, \\ Qingyuan Yang$^{2}$ Adele Bear-Crozier $^{3}$, Michael Pavolonis$^{4}$ and Andrew Tupper $^{3}$\\
$^{1}$Center for Geohazards Studies, \\ University at Buffalo,
Buffalo NY USA\\
$^{2}$Earth Observatory of Singapore, \\ Nanyang Technological
University, Singapore\\
$^3$Bureau of Meteorology, Melbourne, Australia\\
$^4$NOAA Cooperative Institute for Meteorological Satellite Studies \\
University Wisconsin, Madison WI USA}

\begin{document}
\maketitle
\abstract{Volcanic ash clouds often become multilayered and thin with
  distance from the vent.  We explore one mechanism for development of
  this layered structure.  We review data on the characteristics of
  turbulence layering in the free atmosphere, as well as examples of
  observations of layered clouds both near-vent and distally.  We then
  explore dispersion models that explicitly use the observed layered
  structure of atmospheric turbulence.  The results suggest that the
  alternation of turbulent and quiescent atmospheric layers provides
  one mechanism for development of multilayered ash clouds by
  modulating vertical particle motion.  The largest particles,
  generally $> 100 \mu$m, are little affected by turbulence.  For
  particles in which both settling and turbulent diffusion are
  important to vertical motion, mostly in the range of 10-100 $\mu$m,
  the greater turbulence intensity and more rapid turbulent diffusion
  in some layers causes these particles to spend greater time in the
  more turbulent layers, leading to a layering of concentration.  For
  smaller particles, mostly in the submicron range, the more rapid
  diffusion in the turbulent layers causes these particles to ``wash
  out'' quickly.}

\section{Introduction}

Volcanic ash is a multi-billion dollar economic hazard to aviation, as
shown during the 2010 eruptions of Eyjafjallaj{\"o}kull, Iceland
\citep{c94, mazzocchi20102010}. It is also a risk to flight safety,
with hundreds of encounters of varying severity recorded, and several
instances of multiple engine flame-out in flight.  The International
Airways Volcano Watch (IAVW), which seeks to safely separate aircraft
from volcanic ash in flight, relies on detecting areas of ash and
forecasting its future movement \citep{tupper2007facing}.  However,
the forecasting of ash presence and concentration is generally poorly
resolved vertically, although there is some progress in this
direction, e.g., \citep{HEINOLD2012195,
  kristiansen2015stratospheric}. Aircraft flying in a supposedly
ash-contaminated region at a particular altitude may encounter no ash
or significant and potentially damaging amounts, due to the high
degree of ash stratification with altitude. The improved understanding
and forecasting of stratification would assist enormously in managing
the hazard and support the continuing development of the IAVW.

\begin{figure}[!htb]
  \centering
  \includegraphics[width=0.8\textwidth]{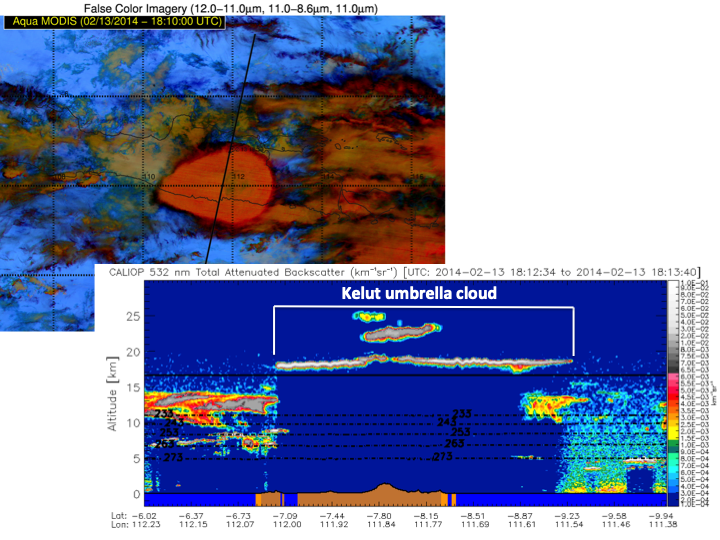}
  \caption{\textbf{(a)} Aqua-MODIS RGB false color image of Java
    capturing Kelud umbrella cloud on 13 February, 2014 (red
    hues). \textbf{(b)} CALIOP nadir LIDAR total attenuated backscatter
    (along track shown in (a) with solid line) showing isotherms
    (kelvin, black dotted lines) and tropopause (black solid
    line). Clouds between 10-15 km on South (right) side under
    umbrella cloud could be of volcanic
    origin. \label{f.example_kelud}}
\end{figure}

\begin{figure}[!htb]
  \centering
  \includegraphics[width=0.8\textwidth]{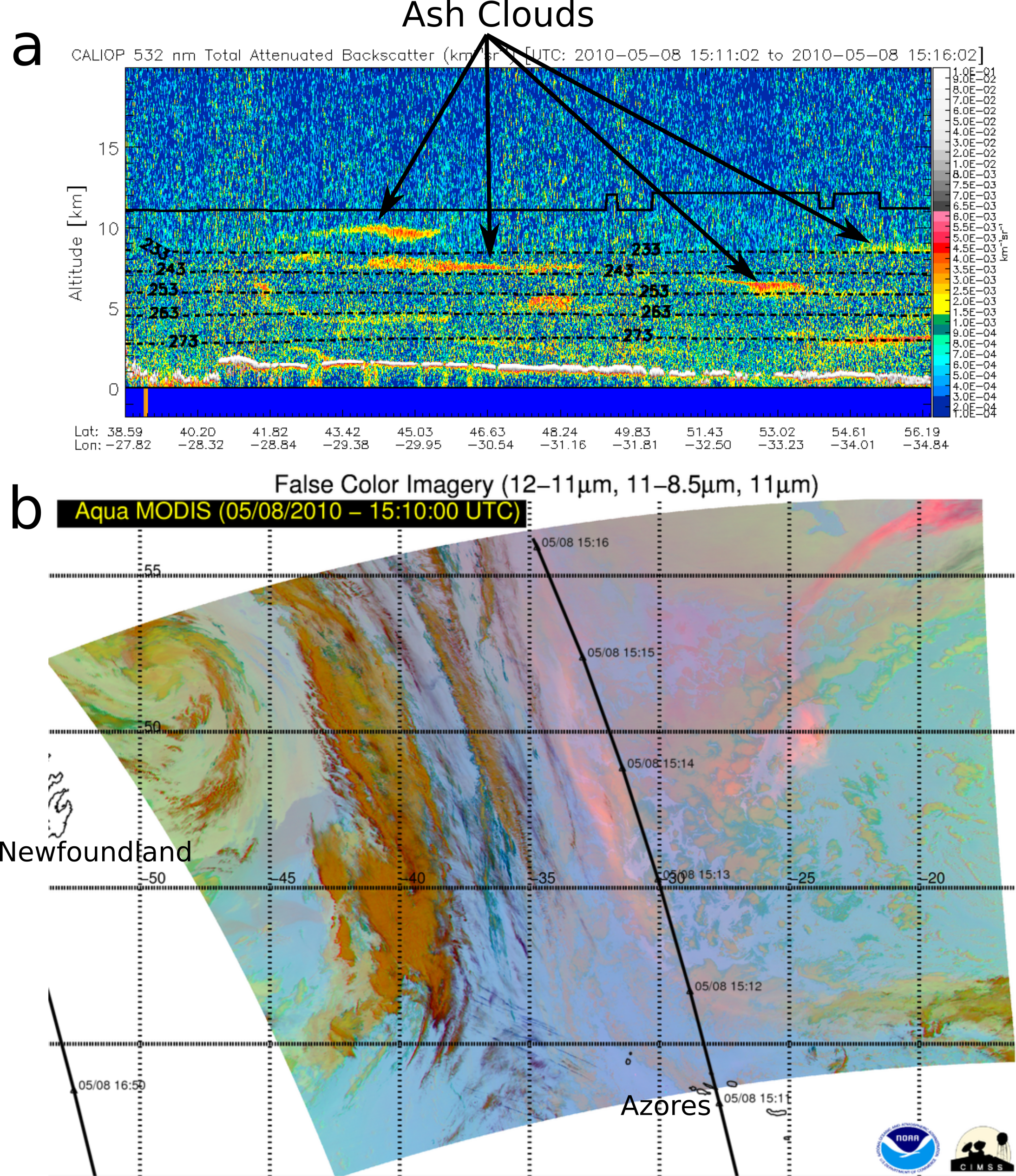}
  \caption{\textbf{(a)} CALIOP nadir LIDAR total attenuated backscatter (along
    track shown in (b)) showing complex layering of Eyjafjallajokull
    ash cloud on 8 May, 2010, isotherms (kelvins, black dotted lines)
    and tropopause (black solid line). \textbf{(b)} Aqua-MODIS RGB false color
    image \citep{PaHeSi13} of North Atlantic
    capturing this ash cloud (pink hues). \label{f.example_eyja}}
\end{figure}

Photography and satellite imagery of numerous volcanic eruptions show
that a single or multiple layered structure is a fundamental aspect of
volcanic cloud development \citep{CaJe13}.  Aerial as well as
ground-based photography have been particularly useful near the
volcano to elucidate this layering.  Lidar backscatter data have been
key in defining the layered structure in distal regions, particularly
those from the CALIOP (Cloud-Aerosol Lidar with Orthogonal
Polarization) instrument (Fig. \ref{f.example_kelud},
\ref{f.example_eyja}).  Volcanic layers can be stratospheric as well as
tropospheric \citep{winker1992preliminary}.


Multiple layers are thought to form and separate by numerous
processes, some unique to volcanic clouds.  The primary volcanic cloud
layer near the vent, such as the volcanic umbrella cloud or anvil
cloud, arises from the driving of hot eruptive gas and ash parcels
outward around their equilibrium level, or neutral buoyancy height
\citep{SpBuCa97}.  As suggested by brightness temperatures over the
surface of near-vent clouds and ground based photography, umbrella
clouds can be solitary, accompanied by a single lower intrusion
resulting from re-entrainment and column-edge downflow \citep{WoKi94,
  Bu98} or accompanied by lower level skirt clouds
\citep{barr1982skirt}, which may or may not contain ash. Ash
accretion, ash re-entrainment and source variability -- injection of
ash at different altitudes with changing eruption rate and wind
fields, cause development of multiple layers \citep{HoSeWo96,
  tupper2004evaluation, thorsteinsson2012high}. Mechanical unmixing of
particulate-laden and gaseous volcanic cloud components, or gas-ash
separation, has been noted as a further cause of volcanic layer
formation \citep{hws96, fero2009simulating}, perhaps enhanced by
gravity current slumping of the particle laden component
\citep{prata2017atmospheric}. Double diffusion and convective sediment
flux to a single \citep{WoKi94, Bu98, HoBuAt99a, HoBuAt99b, CaJe12}
and multiple \citep{CaJe13} levels by descending fingers that intrude
below the level of a major volcanic cloud layer have been observed,
and recreated in the laboratory.

Further from the vent, $\sim >500$ km and hours in transport time,
downward looking satellite sensors are used for cloud tracking, but
these penetrate only partially into the highest layer of an optically
thick cloud, and satellite remote sensing algorithms are most
sensitive to the column integrated ash properties, even when multiple
optically thin layers are present. Although horizontal, planview
resolution can thus be good, satellite data and often volcanic ash
transport and dispersal models (VATDs), used to forecast future
position, have difficulty in reproducing the vertical structure of
distal volcanic clouds \citep{DEVENISH2012152, FOLCH2012165,
  HEINOLD2012195}.  Numerical inversion of VATDs has been used to map
distal cloud positions seen in CALIOP and other data to time-varying
release height from the source vent (\citep{stohl2011determination,
  kristiansen2015stratospheric, ZiLuPo17,
  ZiLuPo18}. \citet{dacre2015volcanic} used a new scheme to calculate
clear-air turbulence (CAT) in the numerical weather prediction (NWP)
model used to drive the NAME VATD model.  Numerous distal clouds from
the \eyja~ eruption were analyzed and thought to be controlled in
thickness by an interaction between wind shear, which acted to thin
layers, and turbulent diffusion, acting to thicken them
\citep{dacre2015volcanic}.

\hypertarget{problem-statement}{%
\subsection{Problem Statement}\label{problem-statement}}

A correct understanding of layer formation and morphology is critical
for any attempt to construct VATD models capable of producing output
consistent with observations of vertical ash distribution, yet the
formation mechanisms of atmospheric ash cloud layers are not fully
understood.  One potential mechanism for their formation and
characteristics is the subject of the present contribution. Our
working hypothesis is that, because particle settling and atmospheric
turbulence act to varying degrees in a layered atmosphere, the
turbulence structure of the atmosphere can cause ash layer formation,
through the process of enhanced suspension in vertically restricted
regions of high turbulence.

\section{Materials and Methods}




\hypertarget{data}{%
\subsection{Background Materials}\label{data}}

\hypertarget{atmosphere}{%
\subsubsection{Atmosphere}\label{atmosphere}}

Ash transport occurs in the large-scale structures of the wind field,
while ash dispersion or spread is caused by the small-scale turbulence
structures and eddies \citep{Bu98, harvey2018multi}.  The free
atmosphere itself, with or without volcanic clouds, generates
small-scale turbulent eddies.  Eddies are generally created in the
troposphere by Rayleigh-Taylor (convective) instabilities associated
with clouds, and can be found 33\% of the time \citep{VaVa98}.  In the
stratosphere, clear-air turbulence (CAT) is primarily thought to be
generated by breaking of upward-propagated gravity waves from the
troposphere or tropopause, i.e., the Kelvin-Helmholtz mechanism,
acting during episodes of vertical wind shear in the horizontal wind
components \citep{barat1982some, nastrom1993coupling,
  sato1995gravity}.  In volcanic clouds near the vent, turbulence is
created by both the Rayleigh-Taylor and Kelvin-Helmholtz mechanisms,
as ash clouds intrude into the atmosphere as gravity currents.
Kelvin-Helmholtz instability is driven by the shear between the
intruding cloud and the atmosphere \citep{britter_simpson_1981}.
Rayleigh- Taylor instability is driven by cloud top convective
instabilities \citep{VaVa98}, convective sedimentation, fingering and
local, eddy scale density reversal \citep{WoKi94, hws96,
  chakraborty2006volcan}.

In the free atmosphere, parameters such as moisture content and
temperature do not change monotonically with height; there are layers
of relatively homogeneous air, separated by regions in which
parameters vary rapidly \citep{VaVa98, ShTrLa12}.  Both the
stratosphere and the troposphere are layered in this way on scales of
$\mathcal{O}[0.1-1km]$ \citep{MaFuYa93, WiLuHa14}, although the layers
in the troposphere tend to be more transient and discontinuous
\citep{gage1980use}.  Information of sufficient resolution in the
vertical direction to discern and characterize the turbulence layered
structure is obtained from airborne measurement campaigns or
rawinsonde balloon releases \citep{dehghan2014comparisons,
  cho2003characterizations, pavelin2002airborne}. Several methods have
been developed to derive turbulence from rawinsonde and other
high-resolution data.  \citet{VaVa98} measured changes in the
refractive index structure parameter for radio waves, as turbulence
causes changes in the refractive index, based on rawinsonde pressure,
temperature, humidity, wind speed and wind direction
data. \citet{clayson2008turbulence} used variations in the potential
temperature profile from an idealized profile to calculate the Thorpe
scale, and derive turbulent dissipation and diffusivity.  These
estimates can be made for single rawinsonde profiles with simple
calculations.  In the troposphere, layers of constant relative
humidity (RH) or mixing ratio, $q$, can act as tracers for high
turbulence \citep{cho2003characterizations}.  This is because as
turbulence intensifies, mixing intensifies, and tracers become more
evenly distributed throughout a layer.  The macroscopic mixing caused
by turbulence acts as a diffusive process, which is characterized by a
(turbulent) eddy diffusivity, $\kappa$.  (Atmospheric moisture is also
important in aiding plume lift, especially in plumes from weak
sources, or at low latitude, where the moisture content is high
\citep{SpBuCa97, TuTeHe09}.  High RH layers might therefore indicate
not only turbulence, but also particularly intense reservoirs of
energy available for plume lift.) As a result of the layering,
large-volume or bulk turbulence, and the eddy diffusivity, $\kappa$,
that expresses its intensity, is highly anisotropic
($\kappa_h \gg \kappa_z$), and only within thin, well-defined layers
is it approximately isotropic
($\kappa_{h, local} \approx \kappa_{z, local}$) \citep{gage1980use}.

There are drawbacks to extrapolating high-resolution or point data to
a regional scale because of spatial and temporal inhomogeneity; the
troposphere is highly transient and spatially variable
\citep{clayson2008turbulence, thouret2000general}; tropospheric
isobaric surfaces are not necessarily parallel to the earth's surface,
especially at fronts and in mountain waves \cite{ShTrLa12}. Fronts are
associated with tropopause folds, non-horizontal isobaric surfaces
separating cold from warm air, and turbulence in folds is generated by
local dynamic and convective instabilities. Mountain waves form as the
density stratified atmosphere flows past the lee side of a mountain or
mountain range.  These waves can break, resulting in local turbulence
concentrated in non-horizontal layers.

\hypertarget{ash-clouds}{%
\subsubsection{Ash Clouds}\label{ash-clouds}}

Data are available from a number of sources on the shape and structure
of both near-vent and distal ash clouds.  In the near-vent region,
data tend to be more limited, due to the lower optical depth.
Nevertheless, cloud brightness temperature (BT) as measured from
nadir-looking geostationary and low-earth orbiting satellites provides
useful information, as the topography of the top of the near-vent
clouds can be quite variable, with a distinct high point or swell
above the central vent that might be many kilometers above the top of
the main umbrella cloud \citep{fero2009simulating}.  In addition,
airborne and ground-based photography and videography have provided
extensive data on the features at the base of the main umbrella or
anvil, and within the underlying cloud layers.  Visible satellite
imaging of the near vent cloud top consistently reveals strong,
well-defined three-dimensional vortex structures above the vent, which
evolve to smooth, somewhat more diffuse structures in the umbrella
cloud \citep{PoBuJo16}.

Although the air can be choked with opaque, diffuse ash bodies that
extend to ground level near vent, and although gravitational
intrusions, such as umbrella clouds, are wedge-shaped by nature and
therefore of variable depth, measurements have been made of the
$\Delta$BT between cloud top and edge of the main cloud.  The results
suggest that umbrella clouds at the vent typically encompass depths of
$\mathcal{O}[5]$ km (Table \ref{t.proximal_depth_ranges}), which makes
a large mass of ash available for transport at multiple levels.  Some
of the lower near-vent clouds also become distal clouds.  In some
cases, it might be possible to find the entire troposphere and even
lower stratosphere charged with ash, or only a distinct layer or two
of $\mathcal{O}[1-10]$ km depth.


\begin{table}[!htb]
  \caption{Examples of main, upper volcanic cloud depth from top to
    cloud edge.  Measured from geostationary imagery in first scene
    after eruption start (photography for Redoubt).  Data from
    \citet{bear2020automated}; and \citet{pouget2013estimationa} and
    \citet{HoSeWo96} (below divider)} \label{t.proximal_depth_ranges}
  \centering
\begin{tabular}{p{1.5in}p{2in}p{1in}p{1in}p{1in}}
  \toprule
  \textbf{Volcano}	& \textbf{Eruption start date}	&
                                                         \textbf{Mean height,
                                                         km ASL} &
                                                                   \textbf{Depth,
                                                                   km}
  & \textbf{No. layers} \\
  \midrule
  Tinakula	& Oct 20, 2017, 2350 UT			& 16.6 & 4.9 &
                                                                       1 \\
  Tinakula	& Oct 20, 2017, 1930 UT			& 15.1 & 3.4 &
                                                                       1 \\
  Rinjani	& Aug 1, 2016, 0345 UT	        	& 5.5	& 4.0
  & 1\\
  Manam		& Jul 31, 2015, 0132 UT			& 13.7 &
                                                                 9.2$\dagger$
  & 2 \\
  Sangeang Api	& May 11, 2014, 0832 UT			& 15.4 &
                                                                 12.0$\dagger$
  & 2 \\
  \textbf{Kelud}		& Feb 13, 2014, 1632 UT			& 15.3 & 2.8 &
                                                                                       1\\
  Manam		& Jan 27, 2005, 1400 UT			& 24.0 & 3.0 &
                                                                       1 \\
  Manam		& Oct 24, 2004, 2325 UT			& 18.5 & 1.5
  & 1\\
  \midrule
  \textbf{Pinatubo} & Jun 15, 1991, 2241 UT & 23.6 & 4.7 & 1\\
  Redoubt & Apr 21, 1990, 1412 UT & 12.0 & 4.9 & 2 \\
  Mount St. Helens & May 18, 1980, 2020 UT & 13.0 & 1.0 & 1 \\
  \bottomrule
  $\dagger$ Depth possibly overestimated. & & \\
\end{tabular}
\end{table}

In the distal region, airborne lidar, EARLINET-AERONET (European
Aerosol Research Lidar NETtwork-AErosol rObotics NETwork) and CALIOP
data typically show much thinner, more discontinuous cloud structures
(Fig. \ref{f.example_kelud}, \ref{f.example_eyja}).  Three separate
tabulations of distal ash cloud layer data for the \eyja ~plume have
been published \citep{jwbdfc96, AnTeSe10, winker2012caliop}, and the
EARLINET data have been summarized and modelled
\citep{dacre2015volcanic}. The data suggest that distal clouds from
this tropospheric eruption were typically 0.3-3 km thick, made up of
2-3 layers, with individual layers of 0.3-1.4 km depth, and maximum
age of 129 h (< 1 week) (Table
\ref{t.depth_ranges}). \citet{dacre2015volcanic} estimated a mean and
standard deviation of the measured depths of $1.2 \pm 0.9$ km.  The
number density of particles in the clouds as they propagated over
Europe generally peaked in the micron to submicron range, while the
mass density peaked near 10 $\mu$m \citep{schumann2011airborne},
although the larger size modes may be underrepresented
\citep{AnTeSe10}.

\citet{vernier2013advanced} using CALIOP data, discerned two or more
well-defined layers in the cloud from Puyehue-Cordon Caulle three
weeks after the eruption. Some of the layers showed fold or
wrap-around structures (Figure 1a in \citet{vernier2013advanced}),
perhaps related to vertical-plane chaotic mixing
\citep{pierce1993chaotic}.  Clouds were up to 3 km thick, with
individual layers of 0.1-2 km depth, in the upper troposphere-lower
stratosphere (UTLS), centered on the tropopause at 8-14 km altitude.

On July 12-13, 1991, 26 days after the last major eruption of
Pinatubo, a lidar flight noted numerous stratospheric layers
\citep{winker1992preliminary}.  The data showed a number of
well-defined layers of 0.5-1 km depth between about 14 and 25 km
altitude (Figure 1 in \citet{winker1992preliminary}).  Along much of
the line of flight there were two layers (22 and 25 km), but in places
there were up to five.  Within the first day, modeling by
\citet{fero2009simulating} is consistent with a mean grain size of 90
$\mu$m in the 22-km layer. Although depolarization ratios in the 25-km
layer were consistent with mostly sulfur aerosols, even after nearly a
month, high backscatter LiDAR depolarization ratios at the base of the
22-km layer were still inexplicably consistent with particles in the
10-100 $\mu$m range \citep{winker1992preliminary}.

In all studies cited above, distal ash layers were horizontal or
tilted relative to the horizon, and had extents in the cross-transport
direction of hundreds of km in the troposphere (\eyja), to thousands
of km in the stratosphere (Puyehue, Pinatubo).

Once the ash propagated far from vent (>11 hr in the case of a large
eruption, e.g., Pinatubo), they no longer retained significant memory
of source conditions \citep{dacre2015volcanic, fero2009simulating}.
However, simple back trajectories of distal ash clouds for
Eyjafjallajökull and Puyehue-Cordon Caull\'e are generally consistent
with theoretically possible cloud heights at the source
\citep{winker2012caliop, vernier2013advanced}. It is not clear that
ash was injected at these altitudes at the source, given uncertainties
in vertical parcel motion and settling speed, or lack of incorporation
thereof in the models \citep{MaPoSi14, vernier2013advanced}.  This
observation suggests that vertical particle motion -- settling -- is
not the sole, or even most, important factor in accounting for
vertical positions of ash clouds in the distal region. A more
consistent feature is thinning, more sharply defining, and multiplying
of the near vent volcanic cloud features.  \citet{dacre2015volcanic}
suggested that increase and decrease in thickness, are likely due to a
balance between vertical wind shear and turbulent diffusion for distal
clouds comprised of mostly micron and submicron particles.

\begin{table}[!htb]
  \caption{Measured distal Eyjafjallajökull cloud depths from CALIOP
    lidar.  Data from \citet{winker2012caliop} (top section),
    \citet{marenco2011airborne} (middle) and
    \citet{schumann2011airborne} (bottom). Note that number of
    layers varies with spatial position.} \label{t.depth_ranges}
  \centering
\begin{tabular}{cccccc}
  \toprule
  \textbf{Date} & \textbf{Cloud}	& \textbf{Height range, km ASL}	& \textbf{
                                                                          Depth, km} &
                                                                                       \textbf{Age,
                                                                                       hr}
  & \textbf{No. layers} \\
  \midrule
  Apr 15 & 20100415 & 1.41–3.23 & 0.51 & $<6$ & -- \\
  Apr 16 & 20100416-a & 3.77–5.50 & 0.58 & 30 & $>1$\\
  Apr 16 & 20100416-b & 1.97–7.27 & 0.67 & 24 & $>1$\\ 
  Apr 17 & 20100417-a & 0.20–6.28 & 0.76 & 42 & 1 \\
  Apr 17 & 20100417-b & 0.05–4.00 & 0.61 & 42 & 1 \\
  Apr 18 & 20100418-a & 3.14–5.59 & 0.81 & 66 & -- \\
  Apr 18 & 20100418-b & 3.75–6.49 & 0.86 & 66 & -- \\
  Apr 19 & 20100419-a & 3.20–5.26 & 1.06 & 71 & -- \\ 
  Apr 19 & 20100419-c & 2.48–3.94 & 0.45 & 30 & -- \\ 
  Apr 19 & 20100419-d & 4.63–5.20 & 0.41 & 114–126 & -- \\
  Apr 20 & 20100420 & 0.05–1.88 & 1.08 & 20–24 & -- \\ 
  \midrule
  May 4 & 20100504 & 2.3–5.5 & 0.5 & -- & 1-2 \\
  May 5 & 20100505 & 2.4–4.5 & 0.9 & -- & 1-2 \\
  May 14 & 20100514 & 5.1–8.1 & 1.1 & -- & 1-3 \\
  May 16 & 20100516 & 3.4–5.5 & 1.2 & -- & 1-3 \\
  May 17 & 20100517 & 3.5–5.6 & 1.3 & -- & 1-3 \\
  May 18 & 20100518 & 2.5–4.9 & 0.9 & -- & 1-3 \\
  \midrule
  \textbf{Apr 19} & 20100419-1 & 3.9-5.6 & 1.7 & 105-111 & $>1$ \\
  Apr 19 & 20100419-2 & 3.5-3.8 & 0.3 & 104-108 & 1 \\
  Apr 19 & 20100419-3 & 3.9-4.2 & 0.3 & 105-108 & 1 \\
  Apr 22 & 20100422-4 & 0.7-5.5 & -- & 49-50 & diffuse \\
  Apr 23 & 20100423-5 & 2.1-3.4 & 1.3 & 40-58 & $>1$ \\
  May 2 & 20100502-6 & 1.6-3.7 & 2.1 & 7.1-12 & $>1$ \\
  May 9 & 20100509-7 & 3.5-4.9 & 1.4 & 97-129 & 1 \\
  May 13 & 20100513-8 & 2.8-5.4 & 0.4-0.7 & 71-78 & 1 tilted \\
  May 16 & 20100516-9 & 3.6-7.0 & 3.4 & 58-66 & $>1$ \\
  May 17 & 20100517-10 & 3.2-6.3 & 3.1 & 66-88 & $>1$ \\
  May 18 & 20100518-11 & 2.8-3.4 & 0.6 & 81-100 & 1 \\
  May 18 & 20100518-12 & 4.0-5.7 & 1.7 & 66-78 & $>1$ \\
  \bottomrule
\end{tabular}
\end{table}

\hypertarget{model}{%
\subsection{Methods}\label{model}}

The atmospheric observations summarized in Section \ref{atmosphere}
show that both the troposphere and stratosphere are layered in
turbulence intensity on a scale of fractions of a kilometer to several
kilometers due to both cloud top turbulence and vertical wind shear
generation.  The ash cloud observations summarized in Section
\ref{ash-clouds} show that more distal, distinct ash clouds evolve
from more proximal, diffuse and thicker clouds.  In VATDs, sub-grid
dispersal in the vertical direction can be described by a single
vertical diffusivity, $\kappa_z$, which may take on the same value as
the horizontal diffusivity, $\kappa_h$; the result is then uniform or
isotropic ash dispersion.  Our goal here is to contrast the physical
behavior of an ash cloud under conditions of isotropic
($\kappa_h = \kappa_z =$ const.) or constant vertical turbulence
($\kappa_z \neq \kappa_z(z)$), and layered turbulence.  We base our
methodology on analytical and numerical models using synthetic
atmospheres with and without multiple turbulent layers separated by
relatively quiescent air.

\subsubsection{Eulerian Analytical Formulation}

In the case of isotropic turbulence, we begin by assuming Cartesian
coordinates, $(x, y, z)$, with velocity components, $(u, v, w)$.  The
three components of the turbulent diffusivity,
$(\kappa_x, \kappa_y, \kappa_z)$ are the same, $\kappa$.  The
concentration of particles in the $i$-size fraction, $C_i$, varies
in time, $t$ and space as:
\begin{equation}
  \frac{\partial C_{i}}{\partial t}+\frac{\partial}{\partial x}\left(u
    C_{i}\right)+\frac{\partial}{\partial y}\left(v
    C_{i}\right)+\frac{\partial}{\partial z}\left(w C_{i}\right) =
  \frac{\partial^{2}}{\partial x^{2}}\left(\kappa C_{i}\right) +
  \frac{\partial^{2}}{\partial y^{2}}\left(\kappa C_{i}\right) + \frac{\partial^{2}}{\partial z^{2}}\left(\kappa C_{i}\right)+\Phi
\end{equation}
where $\Phi$ represents the source/sink function, which in the case of
ash clouds is mostly represented by aggregation and disaggregation of
small particles.  In the present case, such processes are set to zero.
We assume a two-dimensional system with a point-source in time and
space, $w=w_s$, the settling speed, and that, following a streamtube,
the motion of the volcanic cloud can be characterized by a single
downwind coordinate direction $s$ -- for which the axis is everywhere
tangent to the plume centerline, e.g., \citep{w77,hb85} -- and speed
$U$ in that direction. Under these assumptions, the
advection-diffusion equation becomes:
\begin{equation}
  \frac{\partial C_i}{\partial t} + \frac{\partial}{\partial s}\left(U
    C_{i}\right)+\frac{\partial}{\partial z}\left(w_{s} C_{i}\right) = \frac{\partial^{2}}{\partial s^{2}}\left(\kappa C_{i}\right) + \frac{\partial^{2}}{\partial z^{2}}\left(\kappa C_{i}\right)
\end{equation}
with the well-known solution for the impulse initial condition
\citep{Cs80, RoWe02turbulent}:
\begin{equation}
  C_i(s, z, t) = \frac{C_{i0}}{4 \pi \kappa t}\exp\left[-\frac{(s - s_0 - Ut)^2
      +(z - z_0 - w_st)^2}{4 \kappa t}\right]. \label{e.greenfunction}
    \end{equation}
It is reasonably clear that the solution is a Gaussian in $(s, z)$,
in which ash spreads, settles and is blown downwind with time.

The second case is for layered turbulence, in which
$\kappa_z = \kappa_z(z)$, and generally $\kappa_z \ne \kappa_h$.
Since a 1D model is instructive in this system, we investigate
vertical motions alone. In this system, there is a discontinuity in
the diffusive flux at the lower boundary of a turbulent layer because
$\kappa_z$ decreases suddenly, and in this case, the one-dimensional
advection-diffusion equation:
\begin{equation}
\frac{\partial}{\partial t}\left(
  C_{i}\right)+\frac{\partial}{\partial z}\left(w_{s} C_{i}\right) =
\frac{\partial^{2}}{\partial z^{2}}\left(\kappa_z C_{i}\right) \label{e.boundary}
\end{equation}
becomes, in the region of the lower boundary:
\begin{equation}
\frac{\partial}{\partial t}\left(
  C_{i}\right)+\frac{\partial}{\partial z}\left(w_{s} C_{i}\right) \approx 0 \label{e.boundary}
\end{equation}
for $\kappa_{z, upper} \gg \kappa_{z, lower}$.

Thus, a step-like concentration gradient develops at the base of the
upper layer, as some particles are swept back into it rather than
settling across the boundary, hence:
\begin{equation}
C_{i}(t, z)=H(z) C_{i}(t)
\end{equation}
where $H(z)$ is the Heaviside step function, then:
\begin{equation}
\frac{\partial}{\partial t}\left(
  C_{i}\right)+\frac{\partial}{\partial z}\left(w_{s} C_{i}\right) = 
\frac{\partial C_{i}}{\partial t}+w_{s} C_{i} \frac{\partial {H}(z)}{\partial z}=0
\end{equation}

Integrating through the layer depth, $h$, assuming that turbulence is
sufficiently vigorous to homogenize the layer:
\begin{equation}
 \frac{\partial C_{i}}{\partial t} \int_{0}^{h} dz = -w_{s}C_{i} \int_{0}^{h} \delta(z) dz
\end{equation}
we obtain:
\begin{equation}
\frac{d C_{i}}{d t}=-\frac{w_{s}}{ h} C_{i}
\end{equation}
which has solution:
\begin{equation}
  C_i = C_{i0} \exp\left(- \frac{w_s (t-t_0) }{ h}\right)
  \label{e.hazen}
\end{equation}

In a more quiescent layer below the boundary, particles only settle
and are advected downwind, there is little turbulence to enhance
persistence within the layer.  Thus, for systems in which particle
motion is controlled by both turbulent diffusion and settling,
turbulent layers can retain particles longer than do quiescent layers
because of continuing re-entrainment in eddies. Particles fall
relatively rapidly through the quiescent layers because of unhindered
settling, sometimes even enhanced by the effects of convective
sedimentation \citep{HoBuAt99a}, which is not included in the present
model.

\subsubsection{Lagrangian Formulation}\label{lagrangian}

We can gain additional insight by looking at the Lagrangian
formulation of the problem.  In a Lagrangian framework, common to many
ash dispersion models, such as HYSPLIT, movement is given by:
\BE \mathbf{r}_{j+1} = \mathbf{r}_{j} + \mathbf{v} \times \Delta t \EE
where $r$ is the position vector, $v$ is the velocity vector and $j$
is a time index, for which $\Delta t = t_{j+1} - t_j$.  For the
vertical, $z$ component of position:
\BE z_{i+1} = z_{i} + w_{s} \times \Delta t. \EE
If turbulence is added, and assuming no mean vertical flow,
this becomes: \BE z_{i+1} = z_{i} + (w_{s} + w') \times \Delta t \EE
where $w'$ is the vertical component of an instantaneous turbulence
velocity.  Note that the direction (up or down) and magnitude of $w'$
change with time, and therefore effective settling speed, $w_s + w'$,
can be positive or negative (rising parcel) and of widely varying
magnitude, depending on the ratio of $w_s$ to $w'$.  Furthermore, $w'$
can be related to the turbulent diffusion by: \BE \kappa_z = w' L_{o}
\EE where $L_o$ is the length scale of the largest eddies, the Ozmidov
scale, which must be less than the layer thickness, i.e., $L_o < h$,
otherwise layers would be eroded by turbulence.

\subsubsection{Similarity Theory}\label{similarity}

Thus, the velocity scales for turbulence and settling are,
respectively, $\bar{w'}$, the time-mean turbulence velocity, and
$w_s$.  The ratio of these forms the simple dimensionless group,
$\Pi_1$: \BE \Pi_1 = \frac{w_s}{\bar{w'}}. \label{e.pi_1} \EE
Likewise, the timescales for the processes can be used to examine the
conditions under which diffusion or settling dominates.  From Eq
\ref{e.greenfunction}, the timescale of vertical diffusion, $\tau_1$,
through a layer of depth, $h$, is given as
$ \tau_1 = \frac{h^2}{\kappa}$.  From Eq \ref{e.hazen}, the timescale
of settling through the same layer, $\tau_2$, is
$ \tau_2 = \frac{h}{w_s}$.  The ratio of the two timescales indicates
domination of particle transport through the layer by settling or
turbulent diffusion in the vertical direction.  The ratio is given by
the dimensionless group,$\Pi_2$:
\begin{equation}
  \Pi_2 = \frac{\tau_1}{\tau_2} = \frac{h w_s}{\kappa} \label{e.pi_2}
\end{equation}

Potentially important or critical values of particle size and settling
speed, $w_s$, turbulence intensity, $\bar{w'}$, and eddy diffusivity,
$\kappa_z$ are given in Table \ref{t.layering}.  In previous work
describing the generation of ash cloud layering, values of $\kappa$
have ranged from $\kappa_h = 10000$ m$^2$/s and $\kappa_z = 10$
m$^2$/s \citep{fero2009simulating} using the Puff VATD model, to
$10^{-5} \leq \kappa_z \leq 9$, with a mean of 1 m$^2$/s, using the
NAME model \citep{dacre2015volcanic}.  Note that in these models,
larger-scale eddies are assumed to be characterized by the wind field
as described in the NWP model, so that $\kappa$ represents only the
subgrid eddy diffusion.  Thus, the effects of diffusion on eddies of
the scale of the NWP grid might be poorly resolved.
\citet{fero2009simulating} used NWP grids with a resolution of c. 200
km in the horizontal, and 1 km in the vertical, while
\citet{dacre2015volcanic} used an NWP grid with a resolution of c. 50
km in the horizontal, and 100 m in the vertical.  In the Lagrangian
model used in the present contribution, we have explored results for
values of effective $\kappa_z$ ranging from $3 \times 10^{-6}$ to
$6 \times 10^3$ m$^2$/s, while constructing a complete velocity field
following the spectral technique discussed in detail in \citet{Je85}
and \citet{Bu98}.  To obtain turbulence layering in the spectral
model, the Fourier series describing the vertical dispersion is
truncated at different wavenumbers in different layers (Table
\ref{t.layering}).

\begin{table}[!htb]
  \caption{Values of dimensionless groups for different particle sizes
  and layer thicknesses.} \label{t.layering}
  \begin{tabular}{rrrrrr}
\toprule
Diameter & Settling speed, m/s & $\Pi_1$ & & $\Pi_2$ & \\
  $\mu$m & $w_s$, m/s & $\bar{w'} \sim 0.31$ m/s & $\bar{w'} \sim 4.9$ m/s & $\kappa_z = 0.098$
                                                                                     m$^2$/s & $\kappa_z =
                                                                                              5800$ m$^2$/s \\
\midrule
4000 &     5.0e+00 &             1.6e+01 &            1.0e+00 &                     51020.4 &                    8.6e-01 \\
1000 &     3.0e+00 &             9.7e+00 &            6.1e-01 &                     30612.2 &                    5.2e-01 \\
250 &     5.0e-01 &             1.6e+00 &            1.0e-01 &                      5102.0 &                    8.6e-02 \\
100 &     1.0e-01 &             3.2e-01 &            2.0e-02 &                      1020.4 &                    1.7e-02 \\
30 &     1.0e-02 &             3.2e-02 &            2.0e-03 &                       102.0 &                    1.7e-03 \\
10 &     8.0e-04 &             2.6e-03 &            1.6e-04 &                         8.2 &                    1.4e-04 \\
1 &     1.2e-05 &             3.9e-05 &            2.4e-06 &                         0.1 &                    2.1e-06 \\
\bottomrule
  \end{tabular}
  \end{table}

\subsection{Numerical Analysis}
To illustrate effects of turbulence layering on particle settling and
motion, numerical experiments were performed using both Ash3D
\citep{schwaiger2012ash3d} and a Lagrangian model (Section
\ref{lagrangian}).  For a thorough description of the Lagrangian model
and the synthetic atmosphere in which it is run, see \citet{Bu98}.
Experimental parameter values of importance used in the models are
shown in Table \ref{t.experiments}.  In Section \ref{results}, we
explore results from Similarity analysis and numerical solutions based
on the theoretical development.  Numerical solutions are provided for
atmospheres both layered and non-layered with respect to turbulence
(Table \ref{t.experiments}).

\section{Results}\label{results}


In this section, we first investigate expected behavior of ash
particles given a layering in atmospheric turbulence structure using
Similarity Theory.  We then look at numerical results showing what
typical particle vertical speed should be, given a synthetic
atmospheric turbulence layering (Table \ref{t.experiments}).  Output
from several simulations for a synthetic, turbulence-layered
atmosphere using different particle size and settling speed are then
shown to investigate the potential effects of turbulence layering on
particles of different sizes. We then discuss possible turbulence
layering in the atmosphere associated with the climactic 1991 Pinatubo
eruption, and run simulations in the resulting synthetic atmosphere.

\begin{table}[!htb]
  \caption{Simulation parameters. Duration refers to emission from
    vent.  } \label{t.experiments}
  \centering
  \begin{minipage}{\linewidth}
  \begin{tabular}{llll}
    \textbf{Parameter$\downarrow$|Model$\rightarrow$} & \textbf{Test, nonlayered} &
                                                               \textbf{Test,
                                                  layered}
    & 
      \textbf{Pinatubo}  \\
    \midrule
    Simulation type & VATD & Lagrangian & Lagrangian and VATD \\
    Source type & Point & Point & Point \\
    Source height, km & 4.2 & 4.2 - 10 & 24-27 \\
    Particle size, $\mu$m & 10-100 & 1-4000  & 1-1000\\
    Settling speeds tested, m/s & 8.0e-04 -- 1.0e-01 & 1.2e{-05}--1.1 & 1.2e{-05}--1.1 \\
    Amount & 0.01 km$^3$ & 1000 parcels & 1000 parcels \\
    Duration of release & 0.2 hr & Instantaneous & Instantaneous \\
    \midrule
    Turbulent layer heights, km & -- & 2.1-2.7, 3.8-4.2 & 24-25,
                                                          14.5-19 \\
    Turbulent diffusivity, m$^2$/s &  & 0.098-5800\footnotemark  & 0.098-5800 \\
    \midrule
  \end{tabular}
  \footnotetext{$^1$Values are for layers with weaker and stronger turbulence,
  respectively.  Intermediate values also explored.}
  \end{minipage}
\end{table}

We gain insight into expected behavior using results of the Similarity
analysis, and refer to Eqs \ref{e.pi_1} and \ref{e.pi_2} to explore
asymptotic behavior.  In layers for which
$\Pi_1 \parallel \Pi_2 \gg 1$, the turbulent diffusivity is low
relative to the settling speed, the timescale of diffusion is
therefore long, hence motion is controlled by settling.  In layers for
which $\Pi_1 || \Pi_2 \ll 1$, diffusion is rapid relative to settling,
i.e., the timescale of settling is long, hence motion is dominated by
diffusion.  Note also that as layer thickness, $h$, increases, the
timescale of diffusion increases faster than does that for settling,
meaning it becomes more likely the particles will exit a layer by
settling than by diffusion.  For a typical diffusivity of $\kappa = 1$
m$^2$/s in the UTLS \citep{Wi04} at ~10 km altitude, and layer of
depth $h = 1$ km, the critical settling speed, $w_{s, crit}$, dividing
settling from diffusion dominated motion is c. 1 m/s, which would
correspond to a pumice particle of diameter c. 100 $\mu$m at about
2400 kg/cu m (e.g., \citep{ScCoPr05}).

\begin{figure}[!htb]
    \centering
    \includegraphics[width=0.8\textwidth]{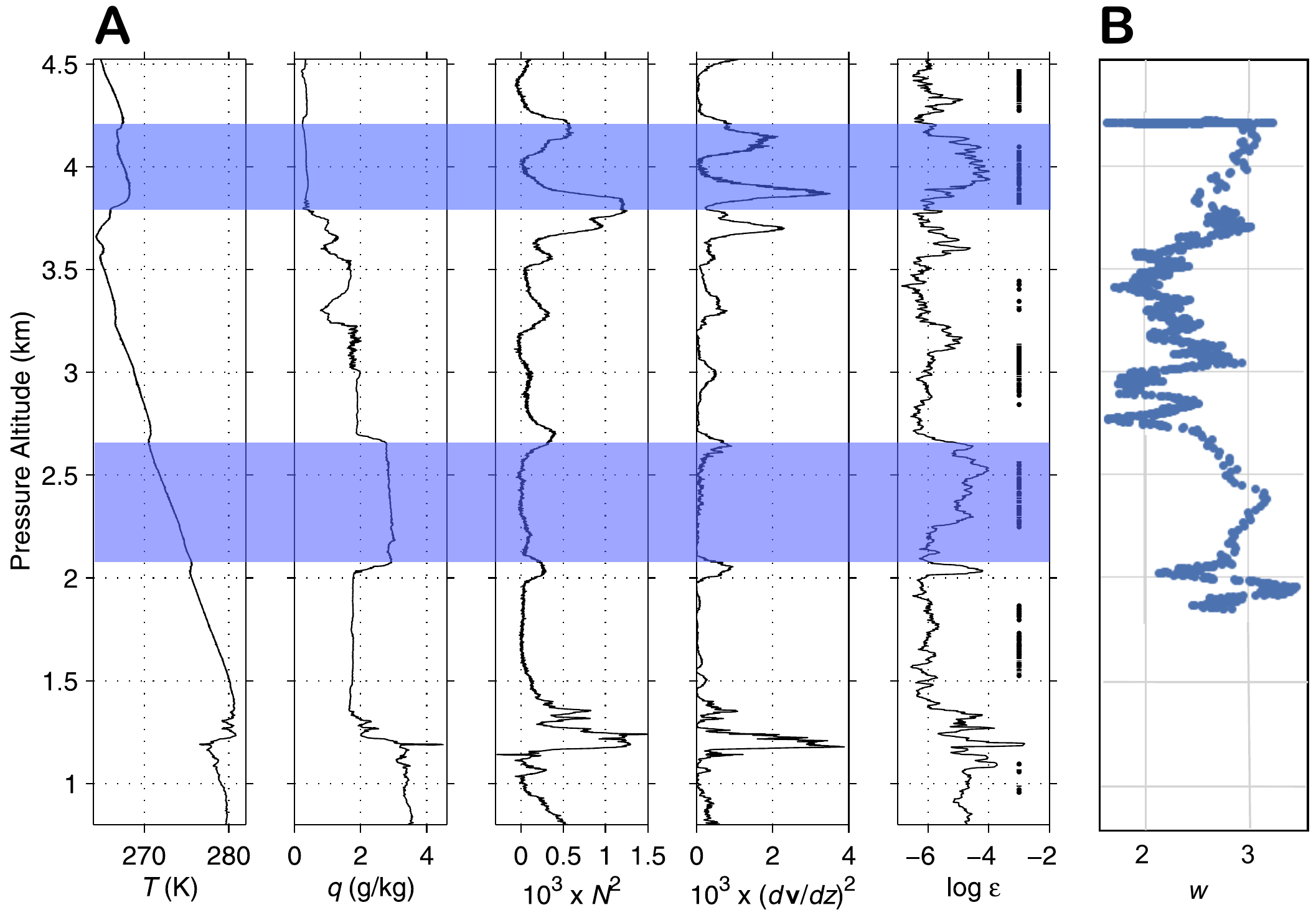}
    \caption{ Turbulent tropospheric layers (shaded) indicated by
      constant mixing ratio ($q$), high turbulent energy dissipation
      rate ($\log \varepsilon$), and bounded by high shear
      ($(dU/dz)^2$). Modified from
      \citet{cho2003characterizations}. Shaded layers are used in
      simplified layer models (Table
      \ref{t.experiments}).  \label{f.layers_Cho_2003}}
   
  \end{figure}

\begin{figure}[!htb]
\centering
  \includegraphics[width=0.7\textwidth]{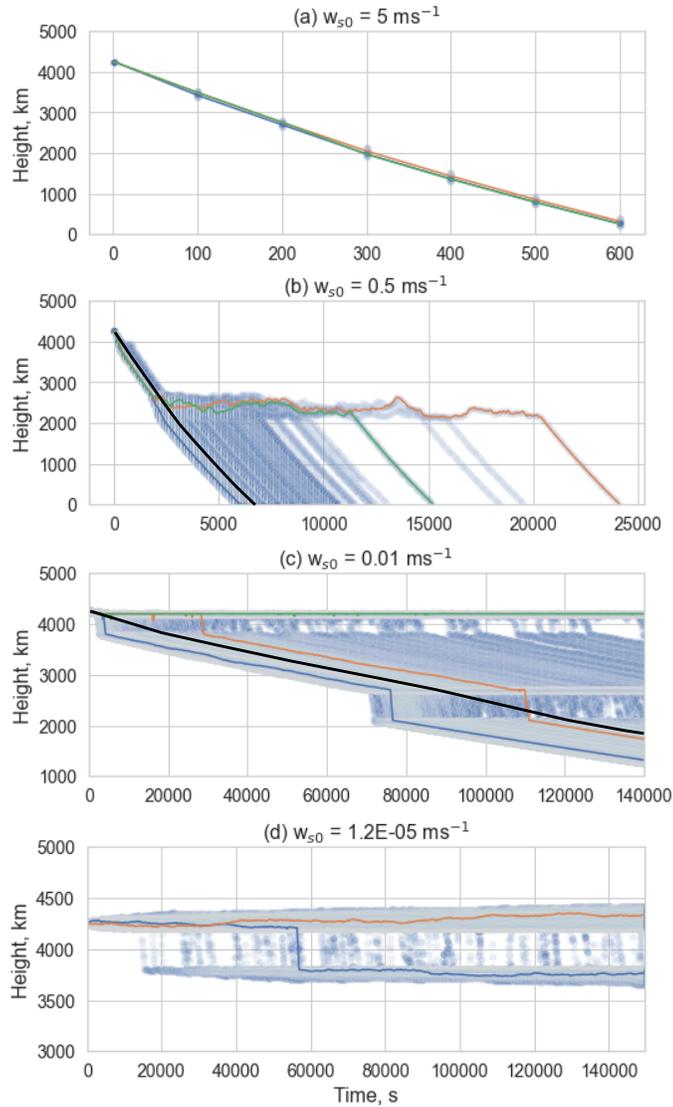}
  \caption{Longitudinal time-section through Lagrangian dispersion
    model showing paths of volcanic particles settling through
    turbulent layers, in which suspension is enhanced. This is output
    based on layers in Fig. \ref{f.layers_Cho_2003}.  See Table
    \ref{t.experiments}.  \label{f.settling}}
  \end{figure}

Numerical results for a layered system (Fig. \ref{f.layers_Cho_2003})
are shown in Figure \ref{f.settling}.  Figure \ref{f.layers_Cho_2003}
is based on the observations of \citet{cho2003characterizations}, who
point out two layers of especially striking turbulence from 2-2.7 km
and 3.8-4.2 km (Fig. \ref{f.layers_Cho_2003}), but do not give
specific values for turbulence intensity.  We therefore apply high
turbulence in these two layers through a simple Lagrangian random
walk, and in other layers, low turbulence.  The largest particles, in
this case $> 1$ mm, are dominated by settling and show little
sensitivity to turbulence (Fig. \ref{f.settling}a).  Intermediate
sized particles, $\sim 10-250 \mu$m (Fig. \ref{f.settling}b, c) in the
more turbulent layers initiate a random walk, being ``stuck'' within
the eddies of the turbulent layer.  These particles are subjected to
turbulence-hindered settling in the more turbulent layers. In these
cases, consider the lower boundary of a layer with strong turbulence.
All the particles there are subject to a random walk. They have a 50\%
percent probability of going up, and a 50\% probability of going down.
Those particles sent above the lower boundary due to turbulence are
sent to a position higher above the boundary than their original
position at the boundary. This will give them a greater chance to
spend a longer time in the turbulent layer, whether or not one
considers settling. Thus, for those layers dominated by the random
walk and dispersion, $\Pi_1 \le 1$, Eq \ref{e.hazen} holds, and the
behavior seen in Figures \ref{f.settling}b, c occurs.  The motion of
the smallest particles, here c. 1 $\mu$m (Fig. \ref{f.settling}d), is
dominated by turbulent diffusion.  The particles spread slowly both
upward and downward in the less turbulent layers, and rapidly in the
more turbulent layers, causing them to have relatively low mean
concentration.

  In contrast, and following from Eq \ref{e.greenfunction}, spread
  from a point source in a VATD model, with isotropic turbulence and a
  wind of constant speed with height, is shown in Figure
  \ref{f.model_xsxn}.  Ash diffuses and progressively spreads from the
  source as the center of mass descends at the settling speed.  Using
  a higher settling speed, the rate at which the center of mass
  descends increases, but the rate at which the particles disperse
  from the center of mass remains constant.  Thus, at any one height
  below the source, particles with a higher settling speed should be
  spread less distance from the source.
                                   
\begin{figure}[!htb]
\centering
  \includegraphics[width=0.5\textwidth]{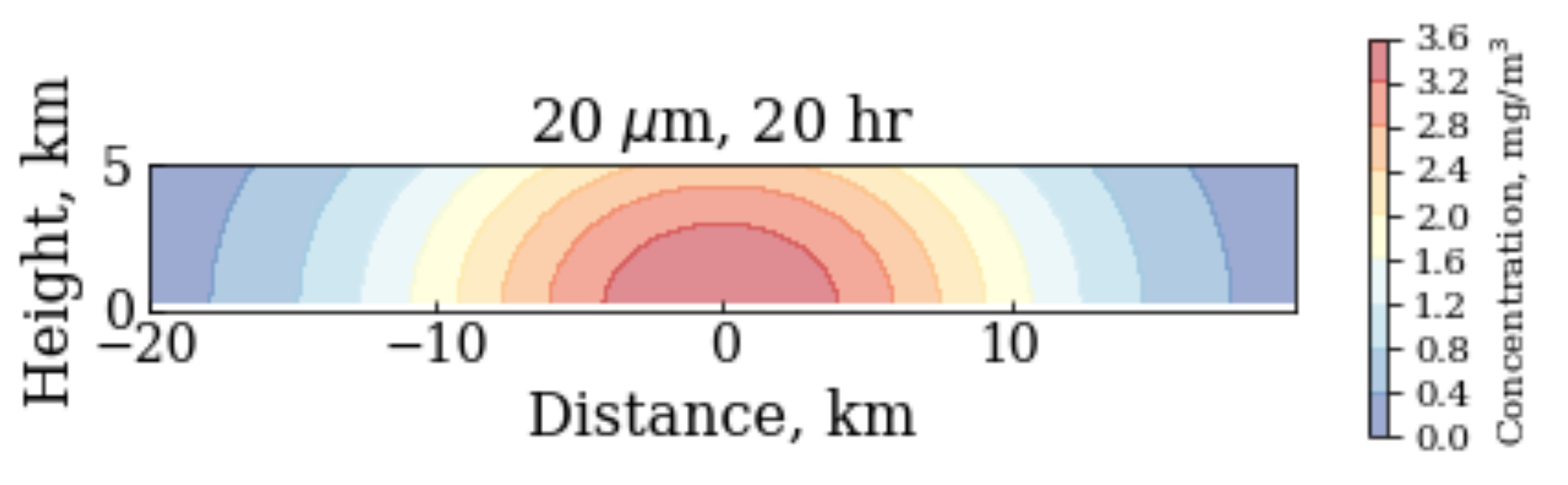}
  \caption{Crosswind section through Ash3D output (Table
    \ref{t.experiments}), with instantaneous source and no wind shear,
    showing advection, settling and isotropic dispersal of $10 \mu$m
    ash.}\label{f.model_xsxn}
  \end{figure}



  \subsection{Application to 1991 Pinatubo eruption}

  There are no direct observations of turbulence values associated
  with the climactic 1991 eruption of Pinatubo.  As noted in Section
  \ref{similarity}, \citet{fero2009simulating} explored values of
  $\kappa_z = 10$ and $\kappa_h = 10^4$ m$^2$/s for the eruption.
  Given understanding of turbulence generation, we can nevertheless
  speculate with some assurance that cloud top turbulence generation
  was probably strong above the Pinatubo umbrella cloud
  \citep{VaVa98}, and that wind shear generated turbulence near the
  tropopause at 16-17 km was probably strong due to the shift from
  southwesterly to northeasterly winds at that altitude
  \citep{fero2009simulating}.  This assessment is consistent with the
  general observation that the most intense level of clear-air
  turbulence development in the lower atmosphere is the tropopause
  \citep{clayson2008turbulence}.  We therefore construct a speculative
  turbulence layering for the eruption based on these considerations.

  The main umbrella cloud of the climactic 1991 Pinatubo eruption was
  centered between 24 to 26 km height and $\sim 3-6$ km thick,
  although ash was injected as low as 17 km and as high as 40
  km\citep{HoSeWo96, fero2009simulating}.  From this initial umbrella
  cloud, the development and separation of layers rapidly evolved into
  two main clouds, the first rich in larger particles and centered at
  the tropopause ($\sim 16-17$ km); the second, a higher cloud rich in
  SO$_2$ and particle that remained centered near the neutral buoyancy
  height of 25 km \citep{fero2009simulating}.  These two main ash
  transport regions persisted, centered around 14-16 and 22-25 km
  \citep{winker1992preliminary} until at least about a month after the
  eruption.  Based on possible/likely generation of stronger
  turbulence by cloud-top convective and wind shear mechanisms,
  respectively, these two regions are given higher values of
  turbulence intensity in layered, Lagrangian model runs (Table
  \ref{t.experiments}).

\begin{figure}[!htb]
\centering
  \includegraphics[width=0.5\textwidth]{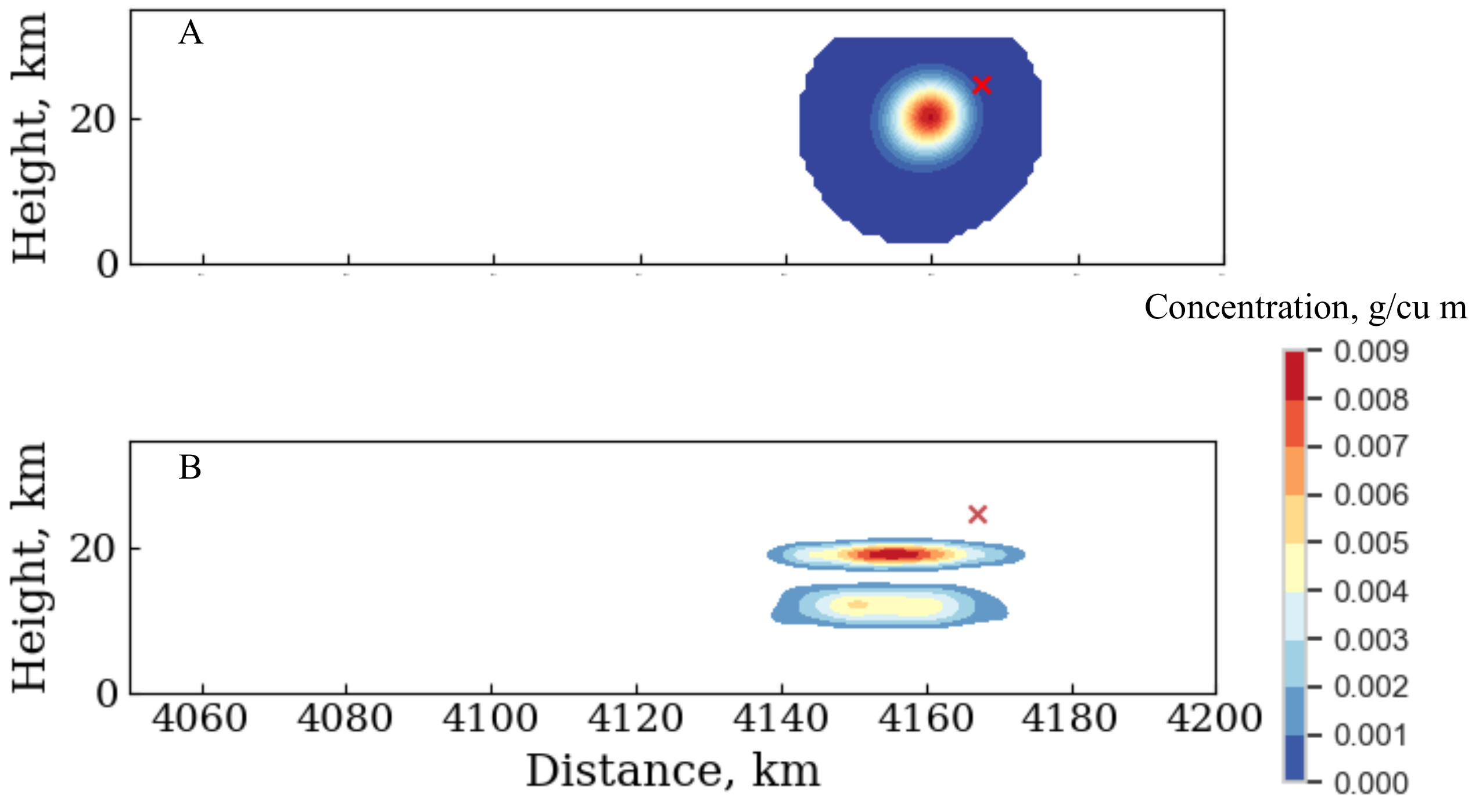}
  \caption{\textbf{(a)} Downwind section through Ash3D output (Table
    \ref{t.experiments}), with instantaneous source and no wind shear,
    showing advection, settling and isotropic dispersal of 1000 $\mu$m
    particles, 50 min after eruption.  \textbf{(b)} Downwind section
    through Lagrangian output (Table \ref{t.experiments}), with
    instantaneous source and no wind shear, showing dispersal of ash
    in layered turbulence.} \label{f.compare_k_z}
  \end{figure}

  With the atmosphere having these speculative intensely turbulent
  layers, we modeled the settling of 1 to 4000 $\mu$m particles from
  the Pinatubo cloud for periods up to a week.  Although particles as
  large as 1000 $\mu$m were affected by the turbulence layering
  (Fig. \ref{f.compare_k_z}), we focus on the results for 10 and 30
  $\mu$m particles at 18hr, as these were dominant in parts of the
  cloud for periods up to nearly a month (Section \ref{ash-clouds})
  \citep{winker1992preliminary}, and as the 18-hr time window is
  particularly important to VAAC ash cloud forecasting.  18 hr is the
  length of the forecast window typically issued in volcanic ash
  advisories (VAA) and shown in volcanic ash graphics (VAG) issued by
  the VAACs.  The results for these particle sizes are consistent with
  observations of cloud heights, particle sizes (Section
  \ref{ash-clouds}) and speculative layering within them
  (Fig. \ref{f.model_xsxn}).  Given particles released from heights
  ranging from 23 to 27 km, after 18hr, 10 $\mu$m particles are
  concentrated in the 24-25 km high-turbulence layers, whereas 30
  $\mu$m particles are mostly spread throughout the 14.5 to 19 km
  high-turbulence layer.  While both boundaries of the 24-25 km layer
  are sharp, $30 \mu$m particles, predominantly in the lower layer,
  are concentrated at and above the 14.5 km lower boundary and around
  the 19 km upper boundary (Fig. \ref{f.profile}).  This concentration
  near the boundaries results from the recirculation of particles from
  the strongly turbulent layer into the overlying less turbulent layer
  near the 19 km boundary (e.g., Fig. \ref{f.settling}d), and settling
  modified behavior near the lower boundary (e.g.,
  Fig. \ref{f.settling}b).  The persistence of the particles in the
  turbulent layers, and more specifically their concentration near the
  lower boundary of the lower layer by turbulence, may explain the
  layering and inexplicable concentration near the base of
  particle-rich layers observed by \citet{winker1992preliminary}.


\begin{figure}[!htb]
\centering
  \includegraphics[width=0.9\textwidth]{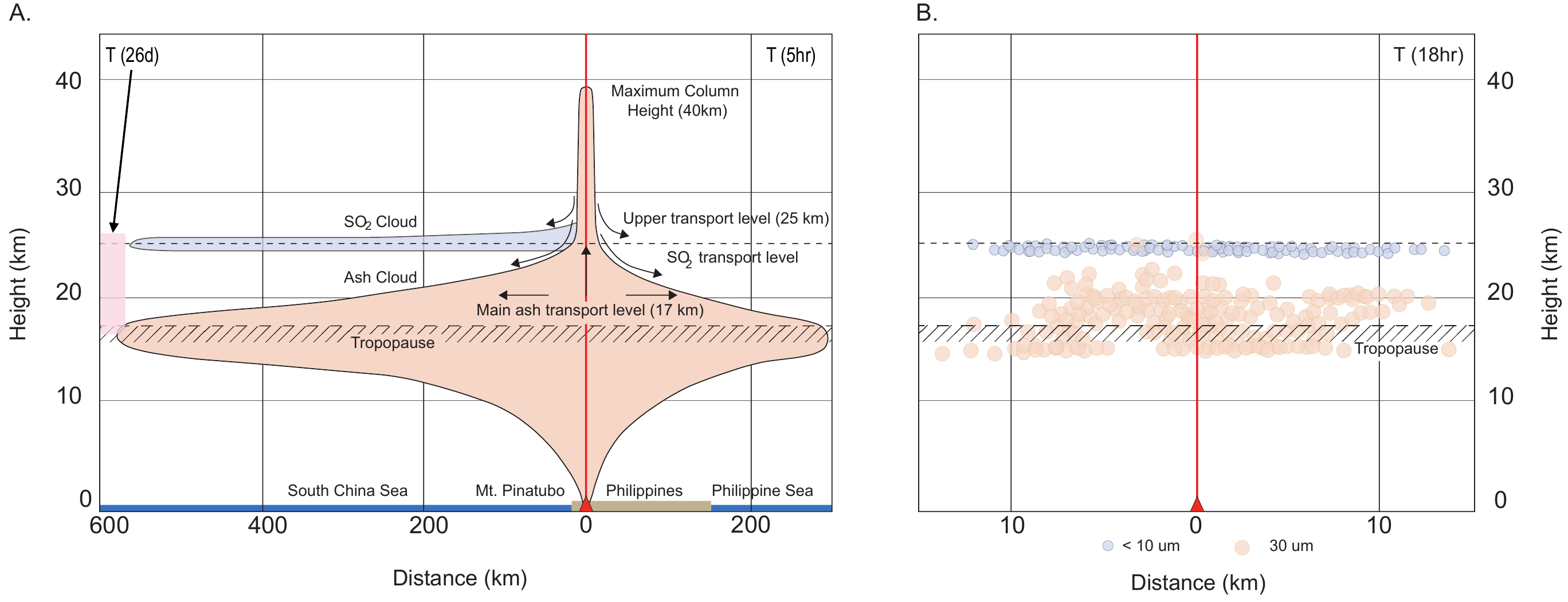}
  \caption{\textbf{(a)} Schematic longitudinal section through
    Pinatubo ash cloud at 5hr, based on
    \citet{fero2009simulating}. Box on left shows range of volcanic
    cloud heights after 26 days according to
    \citet{winker1992preliminary}.  \textbf{(b)} Cross section through
    Lagrangian dispersion model for 10 and 30 $\mu$m particles (Table
    \ref{t.experiments} with turbulence layered atmosphere, showing
    advection, settling and non-isotropic ash dispersal. Red $\times$
    is point of origin for particles, and color gradient is scaled to
    concentration in both. }\label{f.model_xsxn}
\end{figure}

\begin{figure}[!htb]
\centering
  \includegraphics[width=0.5\textwidth]{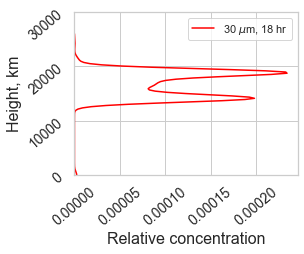}
  \caption{Relative concentration profile of the $30 \mu$m particles
    after 18 hr showing peaks in concentration near top and bottom of
    14.5-19 km turbulent layer.}\label{f.profile}
\end{figure}


\section{Discussion and Conclusions}



In the present work, we have presented data and models on near-vent
and distal volcanic cloud morphology and ash loading.  We have
performed numerical experiments comparing dispersal in an atmosphere
with constant $\kappa$ and variable $\kappa_z$ with height.  The
observational data suggest that distal clouds of depth
$\mathcal{O}$[0.1 to 1] km develop from near-vent clouds of
$\mathcal{O}$[1-10] km depth.  The downwind clouds occur at heights
consistent with the original eruption column heights for both
tropospheric and stratospheric eruptions. The depth range of the
distal layers, being generally less than the near-vent depth range,
and the common stacking of distal layers, suggest that their
development is controlled by atmospheric processes.  The observations
are consistent with the working hypothesis that the layering of the
atmosphere in turbulence intensity, generally causing alternating
suspension and settling dominated behavior of appropriately sized
particles, in the present experiments, those of $\mathcal{O}$[10-100]
$\mu$m, is a cause of distal layer morphology.  Particles larger than
those in this size range are dominated by settling and show little
sensitivity to variations in turbulence intensity, while the motion of
smaller, micron sized particles, is dominated by diffusion, with such
particles being preferentially swept out of more turbulent layers.
The numerical results are consistent with and explain in more detail
those of \citet{dacre2015volcanic} that the motion and layering of
$\mathcal{O}$[1-10] $\mu$m particle clouds is driven by a balance
between wind shear and turbulent diffusion.

In addition to the near-vent volcanic processes modifying ash cloud
layering, atmospheric processes produce distal layered ash clouds,
which are often multilayered due to multiple, alternating layers of
turbulent and quiescent air.  The distal ash layers scale to the depth
of the alternating turbulent and quiescent layers in the atmosphere,
which are $\mathcal{O}[0.1-1]$ km deep. The model outputs
(Fig. \ref{f.model_xsxn}) are not inconsistent with those of VATD and
backtrajectory models, in which homogeneous turbulence can be
associated with layers, given pre-existing ash cloud layering at the
source vent, or wind shear \citep{DEVENISH2012152, FOLCH2012165,
  HEINOLD2012195, winker2012caliop, vernier2013advanced}.  It is
likely that shear together with layering in turbulent diffusivity
controls thickness of distal layers for periods of days to perhaps a
month by hindering settling of particles in the 1-1000 $\mu$m range
\citep{dacre2015volcanic}; note that the wind shear and turbulence
layering processes are not mutually exclusive
\citep{cho2003characterizations}.

The results suggest that to better forecast the position and
morphology of ash clouds for aviation safety and other purposes in
VATDs, the vertical characteristics of the atmosphere need to be
better resolved and characterized, particularly in the 18-hour time
window critical to forecasting in VAAs and VAGs.  Because of the
importance of turbulence and moisture to layer formation, it is
critical that these two parameters especially be estimated well, and
at as high a vertical resolution as possible. Specifically, if
appropriate sonde or other high vertical resolution data are
available, they should be used to calculate heights of potentially
elevated $\kappa$, to be used as deterministic input into VATDs for
the source ash-height injection time history. Thus, it is possible
that the source ash-height injection time history could be determined
\textit{a priori}, rather than found \textit{a posteriori} through
inversion \citep{stohl2011determination,
  kristiansen2015stratospheric}.  Being able to provide such
information \textit{a priori} could have a strong positive impact on
forecasting the future heights and thicknesses of ash clouds for
aviation using VATDs.







\paragraph{Acknowledgments}. Simon Carn is thanked for initial interest and
  encouragement in this work.  Bruce Pitman is thanked for useful
  conversation on this topic.  Larry Mastin gave valuable feedback on
  an earlier version of the ms regarding related work.  Q. Yang was
  funded by the Earth Observatory of Singapore. We thank the creators
  of Ash3D, Pandas, Matplotlib and Seaborn for excellent software
  tools enabling this research.


{The following abbreviations are used in this manuscript:\\

\noindent 
\begin{tabular}{@{}ll}
  AERONET & AErosol RObotic NETwork\\
  BT & Brightness Temperature\\
  CALIOP & Cloud-Aerosol Lidar with Orthogonal Polarization\\
  EARLINET & European Aerosol Research Lidar Network\\
  IAVW & International Airways Volcano Watch\\
  RH & Relative Humidity\\
  UTLS & Upper Troposphere - Lower Stratosphere\\
  VATD & Volcanic Ash Transport and Dispersal
\end{tabular}
}





\bibliographystyle{agu}
\bibliography{ash_strata_clouds}





\end{document}